\documentclass[aps, 
superscriptaddress,
sd,
amsmath,amssymb,
twocolumn, 
twoside]{revtex4-1}

\usepackage{graphicx}
\usepackage[colorlinks = true,
            linkcolor = blue,
            urlcolor  = blue,
            citecolor = blue,
            anchorcolor = blue]{hyperref}

\begin{document}
\newcommand{\transition}{${}^4I_{15/2}-{}^4I_{13/2}{\,}$}
\newcommand{\ercrystal}{Er\textsuperscript{3+}:Y\textsubscript{2}O\textsubscript{3}{ }}
\newcommand{\er}{Er\textsuperscript{3+}}

\title{Dynamic control of Purcell enhanced emission of erbium ions in nanoparticles} 

\author{ Bernardo Casabone}
\altaffiliation{These authors contributed equally to this paper}
\affiliation{ICFO-Institut de Ciencies Fotoniques, The Barcelona Institute of Science and Technology, 08860 Castelldefels, Barcelona, Spain}

\author{Chetan Deshmukh}
\altaffiliation{These authors contributed equally to this paper}
\affiliation{ICFO-Institut de Ciencies Fotoniques, The Barcelona Institute of Science and Technology, 08860 Castelldefels, Barcelona, Spain}

\author{Shuping Liu}
\affiliation{Chimie ParisTech, PSL University, CNRS, Institut de Recherche de Chimie Paris, 75005 Paris, France}
\affiliation{Shenzhen Institute for Quantum Science and Engineering, Southern University of Science and Technology, 518055 Shenzhen, China}

\author{Diana Serrano}
\affiliation{Chimie ParisTech, PSL University, CNRS, Institut de Recherche de Chimie Paris, 75005 Paris, France}

\author{Alban Ferrier}
\affiliation{Chimie ParisTech, PSL University, CNRS, Institut de Recherche de Chimie Paris, 75005 Paris, France}
\affiliation{Facult{\'e} des Sciences et Ing{\'e}nierie,  Sorbonne Universit{\'e}, UFR 933, 75005 Paris, France}
\author{Thomas H{\"u}mmer}
\affiliation{Fakult{\"a}t f{\"u}r Physik, Ludwig-Maximilians-Universit{\"a}t, Schellingstraße 4, 80799 M{\"u}nchen, Germany}

\author{Philippe Goldner}
\affiliation{Chimie ParisTech, PSL University, CNRS, Institut de Recherche de Chimie Paris, 75005 Paris, France}
\author{David Hunger}
\affiliation{Karlsruher Institut f{\"u}r Technologie, Physikalisches Institut, Wolfgang-Gaede-Str. 1, 76131 Karlsruhe, Germany}
\author{Hugues de Riedmatten}
\affiliation{ICFO-Institut de Ciencies Fotoniques, The Barcelona Institute of Science and Technology, 08860 Castelldefels, Barcelona, Spain}
\affiliation{ICREA-Instituci{\'o} Catalana de Recerca i Estudis Ava\c{c}ats, 08015 Barcelona, Spain}


\date{\today}

\begin{abstract}

The interaction of single quantum emitters with an optical cavity \cite{Ritter2012,Wang2019a,Najer2019,nguyen2019} enables the realization of efficient spin-photon interfaces, an essential resource for quantum networks \cite{Kimble2008}. The dynamical  control of the spontaneous emission rate of quantum emitters in cavities has important implications in quantum technologies, e.g. for shaping the emitted photons waveform, for generating quantum entanglement, or for driving coherently the optical transition while preventing photon emission. 
Here we demonstrate the dynamical control of the Purcell enhanced emission of a small ensemble of erbium ions doped into nanoparticles. By embedding the doped nanoparticles into a fully tunable high finesse fiber based optical microcavity, we show that we can tune the cavity on- and out of-resonance by controlling its length with sub-nanometer precision,  on a time scale more than two orders of magnitude faster than the natural lifetime of the erbium ions. This allows us to shape in real time the Purcell enhanced emission of the ions and to achieve full control over the emitted photons' waveforms.
This capability opens prospects for the realization of efficient nanoscale quantum interfaces between solid-state spins and single telecom photons with controllable waveform, and for the realization of quantum gates between rare-earth ion qubits coupled to an optical cavity~\cite{ohlsson2002}.     

\end{abstract}


\maketitle 



Quantum network nodes should be able to store quantum information for long durations, to process it using local quantum gates between qubits, and to exchange this information with distant nodes, ideally using photons at telecommunication wavelengths, via an efficient spin-photon interface \cite{Kimble2008}. Several platforms are currently investigated for the realization of quantum nodes, including  atoms, trapped ions and solid-state systems \cite{Sangouard2011,Northup2014}. Single atoms or solid-state emitters provide a platform for spin-photon interfaces with quantum processing capabilities \cite{Northup2014}. 
The realization of an efficient spin-photon interface is facilitated by the use of an optical cavity in the Purcell regime \cite{Ritter2012,Wang2019a,Najer2019,nguyen2019} which allows channelling the emission from the emitter in the cavity mode, while decreasing the spontaneous emission lifetime. In the presence of dephasing, it also increases the indistinguishability of photons from the emitter. However, a strong reduction of the optical lifetime also reduces the available time to realize quantum gates that rely on dipole-blockade mechanisms achieved by driving the emitter to the excited state~\cite{ohlsson2002}.  
One attractive solution to this problem is to decouple the emitters from the cavity when performing the quantum gates, and coupling it back to emit a single photon at a desired time. The ability to achieve a dynamic modulation of the Purcell factor would therefore be a key ingredient to achieve this type of quantum gates in a high efficiency spin-photon interface, as well as an essential tool to shape the emitted single photon waveform. 
Additionally, the dynamic resonance modulation also enables addressing inhomogeneously broadened ensembles of single emitters as multi-qubit registers.

The control of collective light emission from ensembles of atoms in free space has been demonstrated \cite{Minar2009, Farrera2016}. Experiments towards the dynamic control of emission of the cavity enhanced spontaneous emission rate have been so far mostly performed with semi-conductor quantum dots featuring short optical and spin coherence time \cite{Jin2014,peinke2019} or with Raman schemes with single atoms~\cite{stute2012, morin2019}.
Among solid-state materials, rare earth ion-doped (REI) crystals constitute a promising platform for quantum information processing and networking.  
REI feature exceptional spin coherence time \cite{Heinze2013,Zhong2015a,rancic2018} to store information and narrow optical transitions as spin-photon interface, including at telecom wavelength for erbium ions. They have been used as ensemble based quantum memories \cite{Riedmatten2008,Hedges2010, Clausen2011,Saglamyurek2011,Seri2017}. 
REI possess permanent dipole moments with different values in the ground and excited states, which enables strong dipolar interactions between nearby ions, opening the door to the realization of quantum gates between two or more matter qubits, using a dipole blockade mechanism achieved by exciting one ion~\cite{ohlsson2002}, e.g. between erbium and another ion species in the context of quantum repeaters \cite{KimiaeeAsadi2018}. 

Recently, rare-earth ions have proven to preserve their coherence properties in nanoparticles \cite{Bartholomew2017,Serrano2018,Serrano2019}.  
This facilitates their incorporation into micro cavities to reach strong Purcell enhancement, as necessary for the emission of coherent indistinguishable single photons. Also, it provides the necessary confinement required to isolate close-by single ion qubits (of order of 10 nm distance), as required for dipolar quantum gates. Following the first demonstration in free space~\cite{kolesov2012,kolesov2013,utikal2014}, REIs coupled to nanophotonic cavities have also recently led to the detection and manipulation of single rare-earth ions~\cite{Dibos2018,Zhong2018,Raha2019,Kindem2019}.

In this paper, we demonstrate  the dynamical control of the Purcell enhanced emission of a small ensemble of erbium ions in a single nanoparticle. This is achieved by inserting the nanopaticles into a fully tunable high finesse fiber microcavity at cryogenic temperature,  whose frequency can be tuned at rates more than 100 times faster than the spontaneous emission lifetime of $18.4(7)~$ms of the ions by physically moving the fiber mirror with sub nanometer precision using a piezoelectric device.  This allows us to demonstrate a Purcell factor of 31(2) that can be controlled on a time scale of hundred  microseconds. 
Additionally, we show that we can shape in real time the Purcell enhanced emission of the ions to achieve full control over the emitted photons' waveforms, without perturbing the emitter. 
Our approach opens the door to a solid-state quantum node with the potential of exhibiting quantum computing and communication capabilities all in a single device.

Coupling the emitters to an on-resonance microcavity increases the spontaneous emission rate by the effective Purcell factor
$C = \mathcal{L} \zeta C_0 = \mathcal{L} \zeta \frac{3\lambda^3}{4\pi^2}\frac{Q}{V_m} $,
where $\lambda$ is the emission wavelength, $Q$ the quality factor of the resonator, $V_m$ its mode volume, $\zeta$ the branching ratio of the respective transition and  $\mathcal{L (\delta)} = \frac{(\Delta/2)^2}{\delta^2+(\Delta/2)^2} $ is the detuning factor in case the cavity with linewidth $\Delta$ is detuned by $\delta$ from the emitter's resonance. 
Additionally, the collection efficiency of the cavity mode is given by $\beta = \eta C/(C+1)$ where $\eta$ is the cavity outcoupling efficiency. Thus, fast and near-unity efficiency readout can be achieved for sufficiently large $C$ and $\eta$.

Fiber cavities can achieve high Purcell factors up to $10^4$ \cite{hunger2010}, provide open access to the cavity mode for optimal overlap between the ions and the cavity field \cite{Albrecht2013,Benedikter2017,casabone2018} and are small and lightweight enough to offer fast frequency tuning. 
Nanoparticles are a promising platform to study REI as their small size isolates a mesoscopic ion number, such that the large ratio between the inhomogeneous and homogeneous linewidth can allow one to frequency select single ions \cite{Dibos2018, Zhong2018}. 
Also, their scattering cross section can be small enough such that integration into high-finesse cavities is possible.
In particular, we study \ercrystal nanoparticles with 200 ppm erbium concentration and an average diameter of 150 nm. Y$\textsubscript{2}$O$\textsubscript{3}$ is a host that showed to maintain good crystalline quality in the nanoscale allowing long coherence time for dopants like europium \cite{Bartholomew2017, Serrano2018}  and praseodymium \cite{Serrano2019}. 
For \ercrystal, the optical transition \transition is at $1535~$nm with a relatively large branching ratio $\zeta=0.21$.

\begin{figure}
\includegraphics[width=0.45\textwidth]{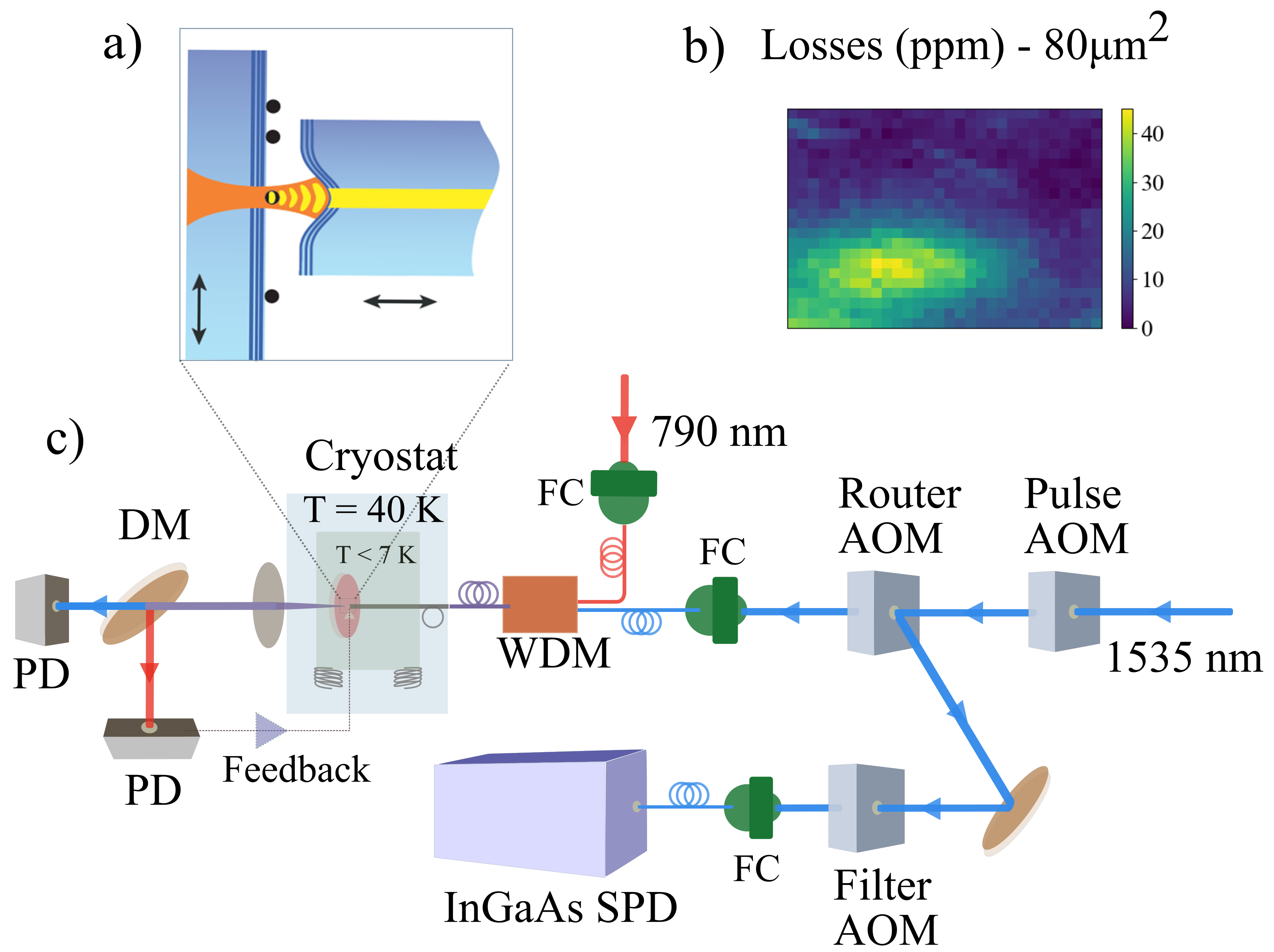}
\caption{
{a) Schematic drawing of the tunable fiber-based microcavity. 
The fiber mirror can be moved in 3D, allowing us to set the length of the cavity and locate a nanoparticle inside the cavity. 
b) Map of scattering loss introduced by a single nanoparticle when scanning the cavity mode across it.
c) The optical setup is described in more details in the SM.  
In summary, a $1535~$nm laser is used to excite the ions and a $790~$nm laser to stabilize the length of the cavity. The two lasers are combined using a wavelength division multiplexer (WDM). 
A set of acousto optic modulators (AOM) are used to create pulses from the CW excitation laser, to route the excitation light to the cavity and the cavity photons to the detector, and to suppress the excitation light to the single photon detector (SPD) by 60 dB (InGaAs, detection efficiency 10 $\%$) during excitation. 
DM is a $780/1535$~nm dichroic mirror, PD are continuous avalanche photo-diodes (APD) for cavity length stabilization and transmission monitoring. 
}
\label{fig1}}
\end{figure}

Our setup is based on a fast and fully tunable, stable, cryogenic-compatible Fabry-Perot microcavity (see Fig.~\ref{fig1}a).
The microcavity is composed of a fiber with a concave structure on the tip, on which a reflective coating is deposited.
The other side of the cavity is a planar mirror with the same reflective coating as the fiber, over which the ions are placed after depositing a thin layer of SiO${}_2$ to ensure maximum coupling between the ions and the cavity electric field (see Supplementary Material (SM)). 
The setup allows to move the fiber around the mirror and localize the nanoparticles, and to set the separation between the fiber and the mirror to form a cavity on resonance with the ions. 
Nanoparticles are located by scattering loss microscopy \cite{mader2015, casabone2018, wind1987}. We use a nanoparticle with a radius of $91(1)$~nm  (see Fig.~\ref{fig1}b) adding an intra-cavity loss of 43(2) ppm and containing close to $11,000$~\er~ions in the measured $C_2$ crystallographic site. 
The empty cavity has a finesse of $2\times10^4$, which is reduced to $1.6 \times 10^4$ in presence of the nanoparticle. 
For cavities as short as $3.5~\mu$m we can expect a maximal Purcell factor $C_{\textrm{max}}=320$, which then reduces to $176$ for a cavity length of $6~\mu$m as used in the work presented here. 

To ensure the highest Purcell factor, a cavity length stability much smaller than $\Delta_{\textrm{FWHM}}$=~$\frac{\lambda}{2 F}\approx 40\,$pm is required. 
We have built a compact and passively stable nano-positioning platform which is robust against the high frequency noise coming from the closed-cycle cryostat such that active stabilization in the low frequency domain, i.e., below $1~$kHz, is enough to stabilize the cavity (see SM).
We use a second laser at $790$~nm to actively stabilize the length of the cavity on an arbitrary point on the transmission fringe. 
The coating is designed to have a finesse of $2,000$ around $790$~nm and such that there is a red mode close by to all $1535$~nm resonances for cavity lengths in the $2\textup{---}20~\mu$m range. 
We were able to stabilize the cavity to a chosen length with residual fluctuations below $30~$pm standard deviation during the whole cycle of the cryostat. 
For the measurements presented here, the stability however is slightly lower ($>100$~pm). 
The temperature of the sample-mirror was below $7$~K. 

In order to perform experiments, we use resonant excitation and detection both via the optical fiber (see Fig.~\ref{fig1}c).
The probability for a photon emitted in the cavity mode to reach the single photon detector is 2.8 $\%$ and the coupling efficiency between the fiber and the cavity mode is calculated to be 55\% (see SM for details). 
The cavity length is set to $6$~$\mu$m including the field penetration depth (one of the shortest accessible modes). 
The optical setup is described in Fig.\ref{fig1}c. 
 
\begin{figure}
\includegraphics[width=0.45\textwidth]{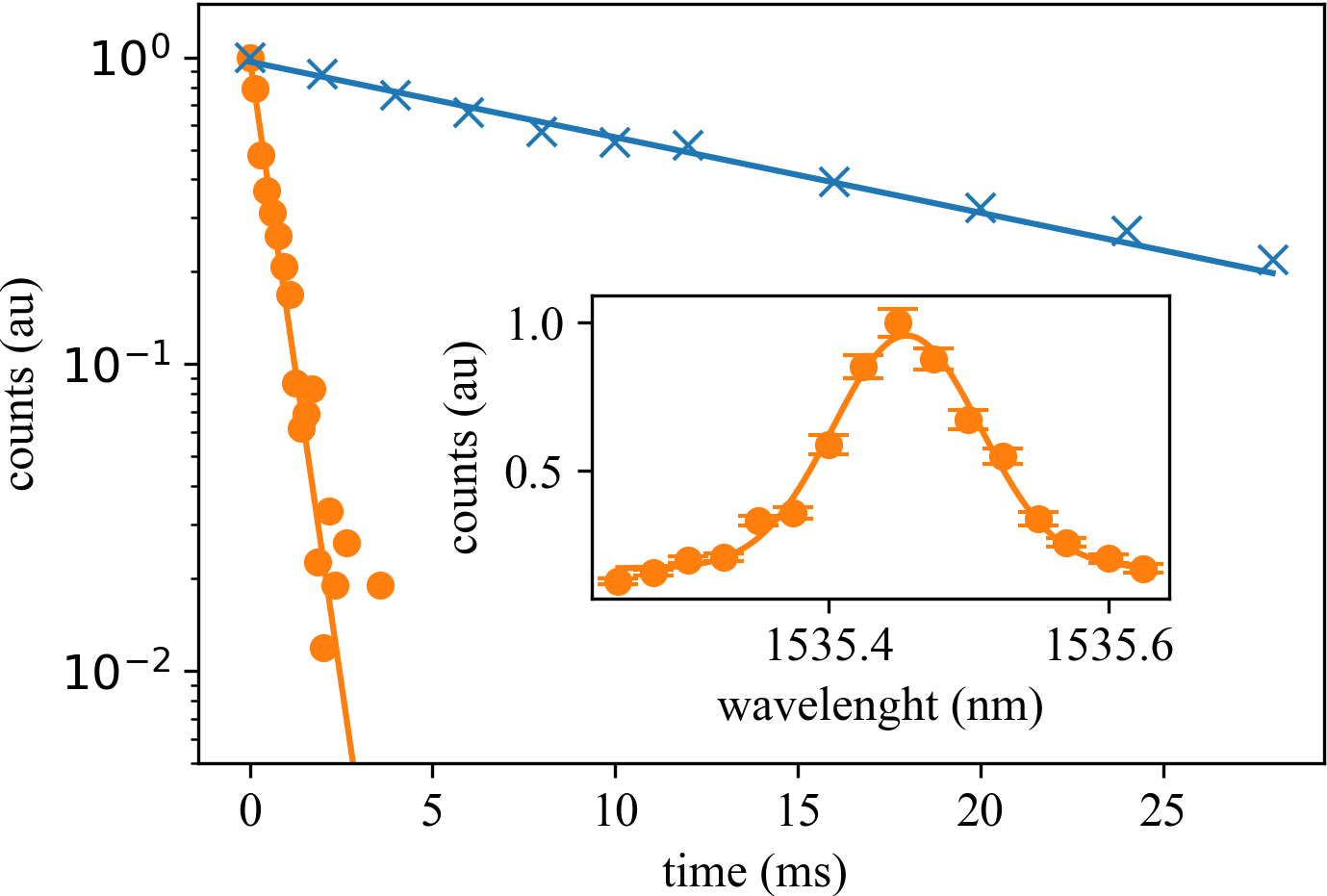}
\caption{
{Crosses: Natural lifetime measured on the studied nanoparticle when it is not coupled to the optical cavity. 
From the fit we extract a natural lifetime $\tau=18.4 (7)$ ms.  
Circles: Lifetime measurement on the studied nanoparticle when it is maximally coupled to the cavity field with $70~$nW input power. 
An exponential decaying fit results in $\tau_c$=0.53 (2)  ms. Bin size is $0.15~$ms.
A background of $15~$Hz was removed in both data sets, calibrated by setting the excitation laser far off resonance. 
Inset: Measurement of the inhomogeneous line of the \transition transition.
The solid line is a Gaussian fit, yielding a full-width-half-maximum linewidth of $15(2)$ GHz.}
\label{fig_lt}}
\end{figure}

We first perform resonant cavity spectroscopy of the \transition transition with an input power (in the input fiber) of $2.4~\mu$W. 
We excite the ions for $300~\mu$s, then wait for $50~\mu$s to ensure the pulse AOM is completely switched off, and collect light for $5$~ms.
The inset of Fig.~\ref{fig_lt} shows the normalized counts in the detection window as a function of the excitation wavelength. 
We fit it with a Gaussian profile and extract a FWHM linewidth of $15 (2)$~GHz centered at $1535.42$~nm.

We then fix the frequency of the laser to be in the center of the line and perform lifetime measurements
 using the same excitation and detection scheme as before but with at input power of $7~$nW.
Fig.~\ref{fig_lt} shows two data sets. 
Circles correspond to the normalized recorded counts
as function of the detection time when the particle is coupled to the cavity. 
From an exponential decaying fit performed over the whole detection window, we extract a lifetime of $\tau_c$=0.53 (2) ms. 
Crosses correspond to the natural lifetime measured in the same nanoparticle when it is not coupled to the optical cavity (see below for a description of the procedure to extract the natural lifetime). 
A fit to the data reveals a natural lifetime $\tau=18.4(7)~$ms which is consistent with an increased excited state lifetime due to the local field suppression in a small nanoparticle \cite{schniepp2002}. 
We then calculate the effective Purcell factor $C=C_0 \zeta = \tau/\tau_c -1$=31(2) which yields $C_0$=31(2)/$\zeta$=147(10) with $\zeta$=0.21 the branching ratio of the transition.  From these high values of $C$ and  $C_0$, we estimate the probability for an ion to emit in the cavity mode $\beta =C/(C+1)$ = 96.9 $\%$, and the cavity enhanced branching ratio $\zeta_{c}= \zeta(C_0+1)/(\zeta C_0+1)= 97.4\%$. 
 
The measured Purcell factor for the ensemble of ions is almost six times smaller than the expected one. 
The reduced value can be attributed to dipole moments randomly oriented with respect to the cavity electric field, the finite extension of the particle with respect of the standing wave \cite{Benedikter2017} and the stability of the cavity which was not optimal for this measurement. 
These effects lead to multi-exponential decay behaviour.
A more detailed analysis reveals that components up to four times faster than $\tau_c$ are present in similar measurements, corresponding to maximal Purcell factors up to $C=150$, consistent with the expected value (see SM). 
The data in Fig.~\ref{fig_lt} corresponds to photons generated by approximately 80 ions due to power broadening, and by decreasing the input power to $300~$pW, a number as low as 10 ions could be detected with a signal-to noise ratio around 5 (see SM). 
High-fidelity single ion detection could be achieved by using more efficient superconducting nanowire single-photon detectors.

\begin{figure*}[tb]
\includegraphics[width=\textwidth]{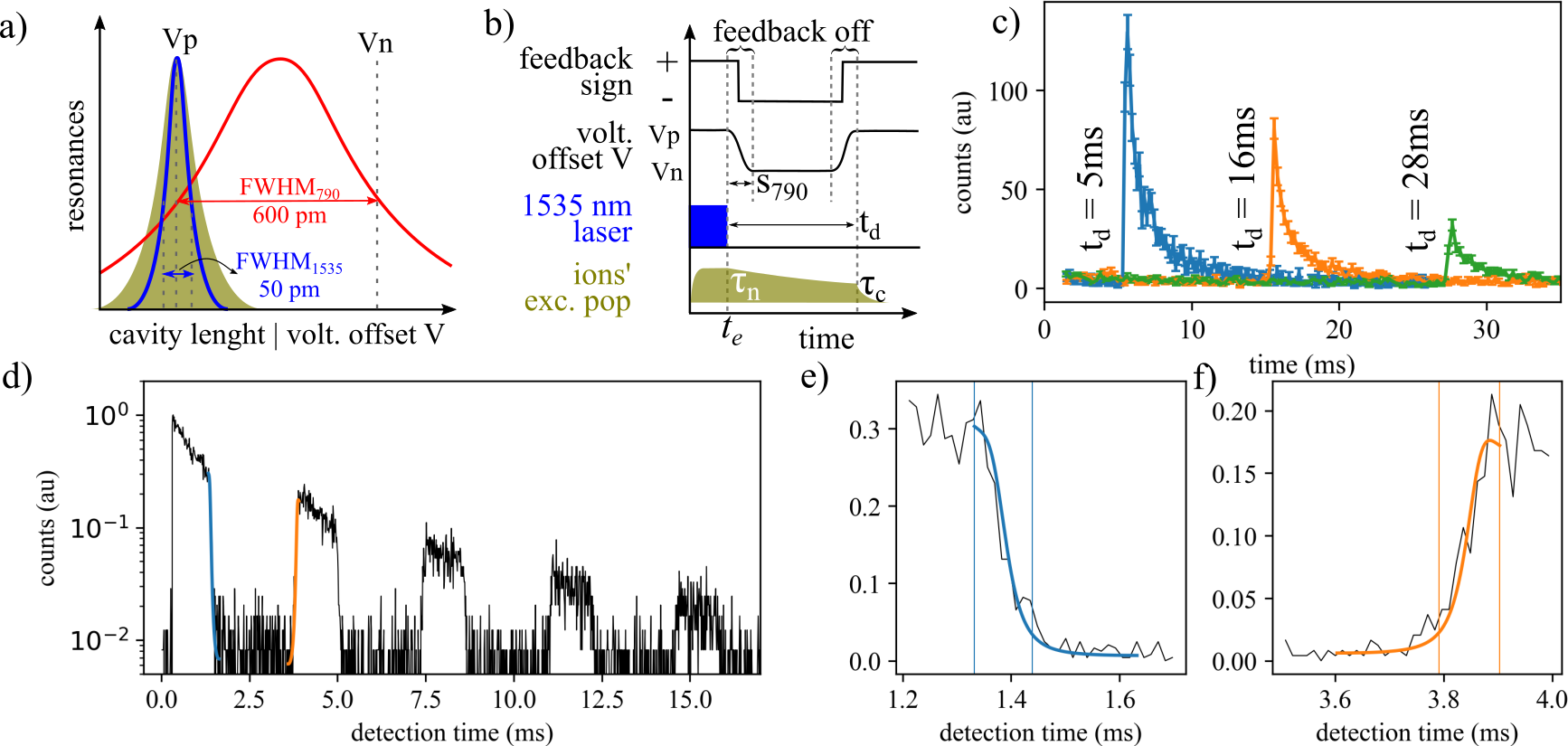}
\caption{
a) Schematic representation of our scheme to stabilize the length of the cavity on/off resonance with the ions.
The cavity length is controlled applying a voltage offset $v$.
For $V_p$, the maximum of the $1535~$nm cavity resonance overlaps with the center of the inhomogeneous line of the ions (solid green area) and with the positive $790~$nm fringe. 
For $V_n$, ions are decoupled from the cavity, which is on resonance with the negative $790~$nm fringe. 
The cavity length can stabilized to the middle of either $790~$nm fringe, for which the feedback sign has to be adequately set.  
b) Sequence used to extract $\tau_n$. 
First, $V_p$ is applied for a time $t_e$ and the $1535~$nm resonance is driven.
At a time $t_e$, the voltage is switched from $V_p$ to $V_n$ in a time $S_{790}$.
The excited state population then decays at $\tau_n$.
After a time $t_e+t_d+S_{790}$, $v_n$ is switched back to $V_p$.
Feedback sign is set to positive (negative) for $V_p$($v_np$), and feedback action is off in the transient stage of the voltage. 
c) Counts recorded as function of time $t$ for $t_d=4, 16$ and $28$ ms. 
d) Counts as function of time when the voltage is alternated between $V_\textrm{p}$ and $V_\textrm{n}$ 5 times at intervals of $1$ ms after excitation of the ions.
Zooms of the falling (e) and rising (f)  edges are also shown, revealing an average transition time of $S_{1535}=106(10)~\mu$s for the falling edge and  $120(10)~\mu$s the the rising (see discussion in main text). 
\label{fig_crs}}
\end{figure*}

We now explain our strategy to tune the cavity resonance in a time scale faster than the spontaneous lifetime. 
We use a second laser at $790~$nm to stabilize the length of the cavity such that the center of the blue-detuned cavity resonance slope overlaps with the maximum of the $1535~$nm ions resonance (see Fig. \ref{fig_crs}a). 
By a fast change of the voltage offset $V$ from $V_p$ to $V_n$ or vice-versa, and by switching the sign of the feedback action, we can stabilize the cavity to either side of the $790~$nm fringe at will (the feedback is off during the transient phase). The voltage is varied as $\sin^2(\frac{\pi t}{2S_{790}})$ and the process happens during a time $S_{790}$ = 300 $\mu s$ for our realization.  
Between the two locking positions, the total fiber displacement is $\Delta_{790}=\frac{\lambda}{2F_{790}}\approx 600~$pm ($F_{790}\approx 700$, reduced due to the particle losses). 
This displacement is 12 times larger than $\Delta_{1535}\approx 50~$pm, such that the detuning between the erbium ions and the 1535 nm cavity is 
$\delta=12\Delta_{1535}$. Using the Lorentzian lineshape of the cavity, one can estimate a maximum reduction of Purcell factor 
$C\propto\mathcal{L (\delta)}\approx 1/570$ and for the fluorescence count rate of $1/634$ \cite{thyrrestrup2013} (see SM).

Next, we show the implementation of this technique and how it can be used to extract the natural lifetime $\tau_n$ of a single nanoparticle while isolating the effect of the cavity from any other lifetime reduction process.
We first set the cavity on resonance with the ions ($V=V_p$) and turn on the excitation  laser at $1535~$nm for $t_e=300~\mu$s (Fig.~\ref{fig_crs}b).
Immediately after, we detune the cavity ($V=V_n$) and start to collect photons.
After a time $t_d$, we set the cavity back on resonance ($V=V_p$) .
Fig.~\ref{fig_crs}c shows the counts as function of time for $t_d=5, 16$ and $28$~ms.
As shown in Fig.~\ref{fig_crs}c, the total counts decreases while increasing $t_d$. 
When ions are decoupled from the cavity, the excited state population decays with the natural lifetime $\tau_n$ and therefore less ions contribute to the detected signal for longer $t_d$. 
In order to extract the natural lifetime, we repeat the measurements for $t_d$ in a range from  $1~\textup{---}~28$  ms and calculate the detected number of photons in the $[ t_d~\textup{---}~t_d + 5 ]$~ms time window (crosses in  Fig.~\ref{fig_lt}). 

Finally, we demonstrate that this technique can be operated with a bandwidth high enough to shape the spontaneous emission of the erbium ions. 
Fig.~\ref{fig_crs}d shows the recorded counts as function of the detection time while the cavity is tuned on (off) resonance for $1~$ms ($2.5~$ms)  5 times. 
Fig.~\ref{fig_crs}e anf Fig.~\ref{fig_crs}f show a zoom-in of the falling and rising edges. The solid lines are a model using the piezo displacement, $S_{790}$ and an effective 1535 nm cavity linewidth to account for cavity instability (see SM). 
Defining the switching time $S_{1535}$ as the time needed to decrease the 
count rate by a factor of 10, we extract from the model $S_{1535} = 106~\mu$s (see SM). 
This value is more than two orders of magnitude shorter than the natural population decay time, and a factor of $5$ shorter than the Purcell enhanced decay time $\tau_c$ of the ionic ensemble.  For future experiments with single erbium ions (or with other ions with shorter lifetime), much higher Purcell factors and consequently much shorter $\tau_c$ will be required. Therefore, much shorter switching times are desirable. Several improvements on our experiments are possible to decrease $S_{1535}$. First, the current measured value is slower than the expected time of $67~\mu$s (see SM), which we attribute to the limited cavity stability. 
Improvements on the cavity stability will therefore directly impact $S_{1535}$. 
Another possibility is to increase the finesse of the cavity at 1535 nm (and therefore decrease $\Delta_{1535}$), as also required for increasing the Purcell factor. Finally, the switching time $S_{790}$ could be decreased by designing a system with higher mechanical eigen frequencies or by  iterative learning algorithms to minimize added noise.
Altogether, we estimate that values of $S_{1535}$ of a few microseconds could be achievable by combining these improvements (see SM). 

In conclusion, we have demonstrated  the dynamical control of the Purcell enhanced emission of a mesoscopic ensemble of erbium ions confined in a nanoparticle embedded in an open fiber based microcavity. By controlling the position of the fiber mirror, we have shown that we can control the cavity resonance and therefore the Purcell factor at a rate more than 100 times faster than the natural decay rate of the ions  with the potential of reaching microseconds switching times. This allowed us to achieve full control on the temporal waveforms of the emitted photons. Combined with single ion addressing, this ability will enable the generation of fully tunable narrowband single photons, and quantum processing using single rare-earth ions. 

\section*{acknowledgements} This project received funding from the European Union's Horizon 2020 research and innovation program under Grant Agreement No. 712721 (NanOQTech), within the Flagship on Quantum Technologies with Grant Agreement No. 820391 (SQUARE), and under the Marie Sk{\l}odowska-Curie Grant Agreement No. 665884. ICFO acknowledges financial support from the Spanish Ministry of Economy and Competitiveness through the ``Severo Ochoa'' program for Centres of Excellence in R{\&}D (SEV-2015-0522), from the Gordon and Betty Moore Foundation through Grant No. GBMF7446 to H. d. R., from Fundaci\'{o} Privada Cellex, from Fundaci\'{o} Mir-Puig, and from Generalitat de Catalunya through the CERCA program.

\clearpage

\renewcommand{\theequation}{SM.~\arabic{equation}}
\renewcommand{\figurename}{Fig. SM.}

\onecolumngrid \section*{Supplementary material: Dynamic control of Purcell enhanced emission of erbium ions in nanoparticles}

\twocolumngrid

Here we discuss aspects relevant to the experiment which have not been addressed in the main text.
First, in section \ref{sec_cav_par}, we present more details of the optical cavity and the optical setup. 
Then, in section \ref{sec_loc}, we describe the technique we use to localize the nanoparticles and to ensure maximal coupling to the cavity electric field.
We calculate the size and estimate the number of erbium ions present in the nanoparticle. 
In section \ref{sec_cav_stab} we introduce our cavity stabilization mechanism and estimate the cavity stability.
Then, in section \ref{sec_cab_res} we describe the method we derived to switch the cavity resonance at will, characterize the speed of the switch and finally discuss limitations and possible improvements to decrease switching time. 
In section  \ref{sec_mul} we show measurements of the reduced excited state lifetime and analyze the decay as function of the time window.
We present evidence of multi-exponential decay behavior and discuss possible causes. We then identify the fastest and the average Purcell factor.
Finally, in the section \ref{sec_det_sig} we specify the efficiencies of our setup and discuss the feasibility of our current apparatus to detect a single erbium ion and possibles modifications to reach high fidelity detection.

\section{Cavity and optical setup}\label{sec_cav_par}
The microcavity is composed of a fiber with a concave structure on the tip on which a reflective coating is fabricated. The other side of the cavity is a planar mirror with the same reflective coating as the fiber. To ensure maximum coupling between the ions and the cavity electric field, we add a SiO${}_2$ layer of $245~$nm thickness.
The thickness of the layer is chosen such that the maximum of the electric field is $40~$nm above the  SiO${}_2$-air interface, where nanoparticles are deposited. 
The  transmission  of the fiber is $T_f = 100~$ppm, while for the mirrors $T_m \approx 2 \times T_f$ due the presence of the SiO${}_2$ spacer thus leading to a maximum finesse of $F_{\textrm{max}} \approx 20,000$ and a total field penetration depth of $2~\mu$m. 
The radius of curvature of the concave structure is $r_{oc} \approx 50~\mu$m and the depth of the structure is close to $p_{d} \approx 1.5~\mu$m

The optical setup is shown in Fig~1c. 
It allows us to perform resonance excitation and detection both via the optical fiber.
$1535~$nm laser is used to excite the ions and $790~$nm laser to stabilize the length of the cavity.
An acousto-optic modulator (AOM) in a double pass configuration is used to `pulse' the excitation laser.
A single pass AOM operates as excitation/detection `router'.
During excitation, the router AOM is off and the excitation light is directed to the cavity while most of the reflected light is directed back to the same channel.  
During detection, the router AOM is on and the spontaneous emission from the ions is deflected towards an InGaAs single photon counter (detection efficiency 10 $\%$). 
A `filter' double pass AOM after the router is used to add additionally protection ($60$~dB) to the single photon detector during excitation.  
A wavelength-division multiplexing (WDM) and dichroic mirrors (DM) both for $780/1535$ are used for merging the light directed onto the cavity and to separate the transmitted light. 
The transmitted light is then directed to continuous APDs (PD) for cavity length stabilization and transmission monitor.

\section{Localizing nanoparticles}\label{sec_loc}
To localize a nanoparticle, we use scattering loss spectroscopy~\cite{mader2015_2}. 
While scanning the fiber cavity, we monitor the cavity transmission which is given by 
\begin{equation}
T_c=\frac{4T_fT_m}{(T_m + T_f + 2B)^2 }
\end{equation}
where 
$T_f$ and $T_m$ are the transmission losses of the fiber and the mirror, $B=4\sigma/\pi w_0^2 $ the additional losses due scattering introduced by the nanoparticle with $w_0$ the cavity mode waist and $\sigma$ the scattering cross section. 
The latter can be calculated as 
\begin{equation}
\sigma=\left(\frac{2\pi}{\lambda}\right)^4\frac{\alpha^2}{6\pi}
\end{equation}
where the polarizability of the nanoparticle is given by~\cite{wind1987_2}
\begin{equation}
\alpha=3\epsilon_0 V \frac{n^2-n^2_{air}}{n^2+2n^2_{air}}
\end{equation}
with $n$ and $n_{air}$ the refractive indices of the nanoparticle and surrounding medium, $\epsilon_0$ the vacuum dielectric constant, and $V=4/3\pi r^3$ the volume of the nanoparticle.

We look for nanoparticles which are big enough to introduce a visible scattering signal, but small enough to maintain the out-coupling efficiency 
\begin{equation}
\eta_{out}=\frac{T_{out}}{(T_f + T_m + 2 B)} \label{eta_out}
\end{equation}
at a high level, where $T_{out}$ is the out-coupling channel, that is, $T_f$ or $T_m$.
Fig.~1b (main text) shows a map of the losses that the studied nanoparticle introduces to the cavity at $1535~$nm.
For $T_f=T_m / 2 = 100~$ppm, a peak loss $B\approx 43(2)~$ppm is measured when the nanoparticle is well aligned to the cavity mode. 
For results presented here, the cavity length is $6~\mu$m, which includes the field penetration depth. 
For a radius of curvature of the structure of the fiber of $50~\mu$m, we calculate the beam waist $w_0=2.9~\mu$m. 
With $n=1.9317$ the refractive index of Y${}_2$O${}_3$, we infer a nanoparticle radius of $90.5(1.0)$~nm.
Finally, for an erbium doping concentration of $200~$ppm, the total number of erbium ions in the crystallographic
site of C2 symmetry is close to $11,000$.

\section{Cavity stability}\label{sec_cav_stab}

\begin{figure*}
\centering{\includegraphics[width=2\columnwidth]{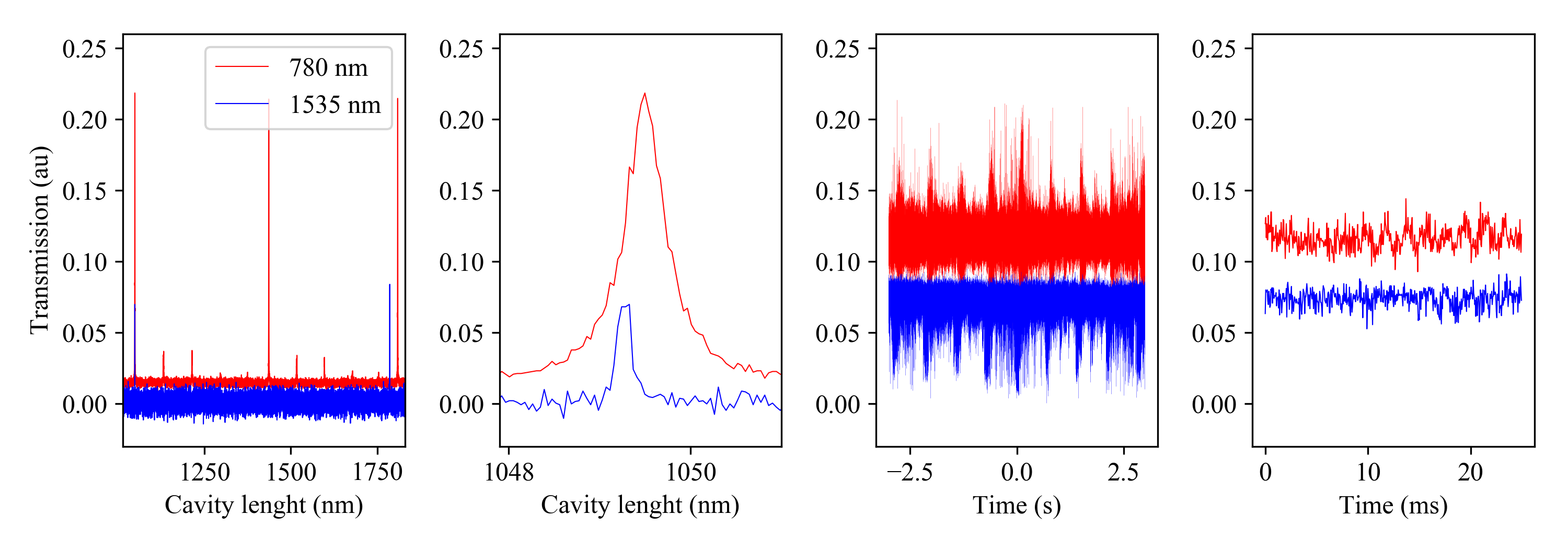}}
\caption{
a) $780~$nm and $1535~$nm cavity transmission as function of cavity length. 
For this measurement, the cavity finesse at $780~$nm and $1535~$nm was $700$  and $5,000$ respectively.
b) Zoom in of (a) on the fringe used to stabilize the cavity length. 
b) Transmission during $5$ seconds while the cryostat is on. 
The cavity length is actively stabilized to the side fringe of the $780~$nm transmission while it is on resonance at $1535~$nm. 
By calculating the standard deviation of the $790~$nm transmission, we infer a cavity stability of $30~$pm
c) Zoom in of (c) on 30 milliseconds over which the cavity stability is the highest.	
}
\label{locking}
\end{figure*}

The cavity is placed on a compact and passively stable nano-positioning platform, which is based on the cryogenic cavity design currently commercialized as qlibri cavity platform, and is robust against the high frequency noise coming from the closed-cycle cryostat. In order to stabilize the cavity length at $1535~$nm, which is the ion's resonance, we make use of a second stop band at $790~$nm.
The coating is designed such that for every $1535~$nm mode, there is a close by $790$ mode for cavity length stabilization for lengths in the $2$ to $20~\mu$m range.  
Fig. SM. \ref{locking}a shows a transmission scan at $1535~$nm and $790~$nm while scanning the cavity length by close to $1~\mu$m.  

To stabilize the length of the cavity, we fine tune the wavelength of the $790~$nm laser such that the maximum of the $1535~$ nm transmission peak overlaps with the middle of the 790 fringe (see Fig. SM. \ref{locking}b).
Then, the transmission of the red laser is used to monitor cavity drifts and a feedback is applied to the piezoelectric crystal to keep this signal at a constant level (side of fringe lock).  
Fig. SM. \ref{locking}c shows the transmission of both lasers as function of the time while the feedback system is on for the whole cryostat cycle.  

For this particular measurements, the finesse $F$ of the red cavity is $700$ and the one of the telecom cavity is $5,000$. 
We estimate the cavity stability as 
\begin{equation}
\frac{ \Delta_{790}} {V_{\textrm{pp}}} \times V_{\textrm{std}}\approx 30~\textrm{pm} \label{cav_stab}
\end{equation}
where $ \Delta_{790} = \frac{\lambda/2}{F}$ is the FWHM of the fringe in units of distance, $V_{\textrm{pp}}$ and $V_{\textrm{std}}$ are the peak-to-peak and the standard deviation of the the voltage measured for the $790~$nm fringe.

For future reference, we also compare the average transmission of the $1535~$nm fringe with the maximum value.
For this measurement, the normalized average transmission is $0.80$.
We now calculate a peak-to-peak displacement $n$ in units of $\Delta_{1535}$ such that the integral over the Lorentzian spectral line matches the average normalized transmission, that is, 
\begin{equation}
\int_{-n/2}^{n/2 }
\frac{1}{2} \frac{1/2}{x^2+(1/2)^2} \frac{1}{x} \, \mathrm{d}x  = 0.8
\label{p2p}
\end{equation}
From the equation we extract $n\approx 1$, corresponding to a peak-to-peak displacement of $n \times \Delta_{1535}=150~$pm.

In Fig.~SM.~\ref{locking}c, we can identify time intervals which show significantly smaller signal dispersion, this interval corresponds to the quietest part of the cryostat cycle, which is close to $500~$ms.
Fig.~SM.~\ref{locking}d corresponds to a zoom in of $30$~milliseconds over which the cavity stability is the highest, for which we measured a cavity stability as defined in \ref{cav_stab} of $20~$pm.

As seen in this section, our positioner is stable enough to keep the cavity on resonance with the ions during the whole cycle. 
We note that the cavity stability for the data shown in the main text and in the next sections is twice worse.
The reason for that is not a fundamental limitation. 
The high stability reported in this section was recorded in the first assembly of the positioner, while for the data shown in the next sections the positioner was re assembled several times.
A possibility exists that during the subsequent assemblies a component was not optimally placed resulting in a degraded stability.

\section{Cavity resonance switching}\label{sec_cab_res}

As discussed in the main text, we switch the cavity resonance by a fast change of a voltage offset on a piezoelectric and by stabilizing the cavity to either side of the $790~$nm cavity fringe at will. 
Between the two sides of the fringe, the total cavity length displacement is $\Delta_{790}=\textrm{FWHM}_{790}$.
The $1535$~nm resonance with  linewidth $\Delta$ overlaps with the middle of one side of the $790~$nm fringe (see Fig.~3a and description from main text), thus when the cavity is stabilized to the opposite side, erbium ions are detuned from the cavity.

Now, we characterize the reduction of the Purcell factor and the countrate during the time the cavity resonance is switched.
The Purcell factor as function of the detuning $\delta$ from the cavity resonance $\Delta$ is then given by 
\begin{equation}
C({\delta})=\mathcal{L (\delta)} \cdot C, 
\end{equation}
where 
\begin{equation}
\mathcal{L (\delta)} = \frac{(\Delta/2)^2}{\delta^2+(\Delta/2)^2}, 
\end{equation}
is the normalized Lorentzian spectral line of the cavity.

The countrate in the cavity mode as function of the time is given by \cite{thyrrestrup2013_2}
\begin{equation}
r(t)= \frac{C(t)}{\tau_n} \cdot \mathcal{N}(t)
\label{model}
\end{equation}
where $\tau_n$ is the natural lifetime, $\mathcal{N}(t)$ is the number of ions in the excited state and $C(t)$ is the Purcell factor.
The number of ions in the excited state can be calculated as 
\begin{equation}
\mathcal{N}(t)=\mathcal{N}_0 e^{- \frac{t}{\tau_n} -\int_{0}^{t}\frac{C(t')}{\tau_n} dt' }
\label{eqnt}
\end{equation}
where $\mathcal{N}_0$ is the number of ions in the excited state before the cavity frequency is shifted.
The first term in the exponent of Eq. \ref{eqnt} takes into account the decrease in population due to emission in free space, while the second term doest it in the cavity mode.
Eq. \ref{eqnt} only take into account the modification in the density of states and neglects light-matter dynamics \cite{thyrrestrup2013_2}.

In order to minimize coupling of mechanical noise in the cavity, the detuning as function of time, that is, the voltage applied to the piezoelectric transducer, is given by 
\begin{equation}
\delta(t) = \Delta_{790} \cdot \sin^2\left(\frac{\pi t}{2S_{790}}\right )
\end{equation}
where $S_{790}$ is the switching time between the two locking points.

\begin{figure}
\includegraphics[width=0.45\textwidth]{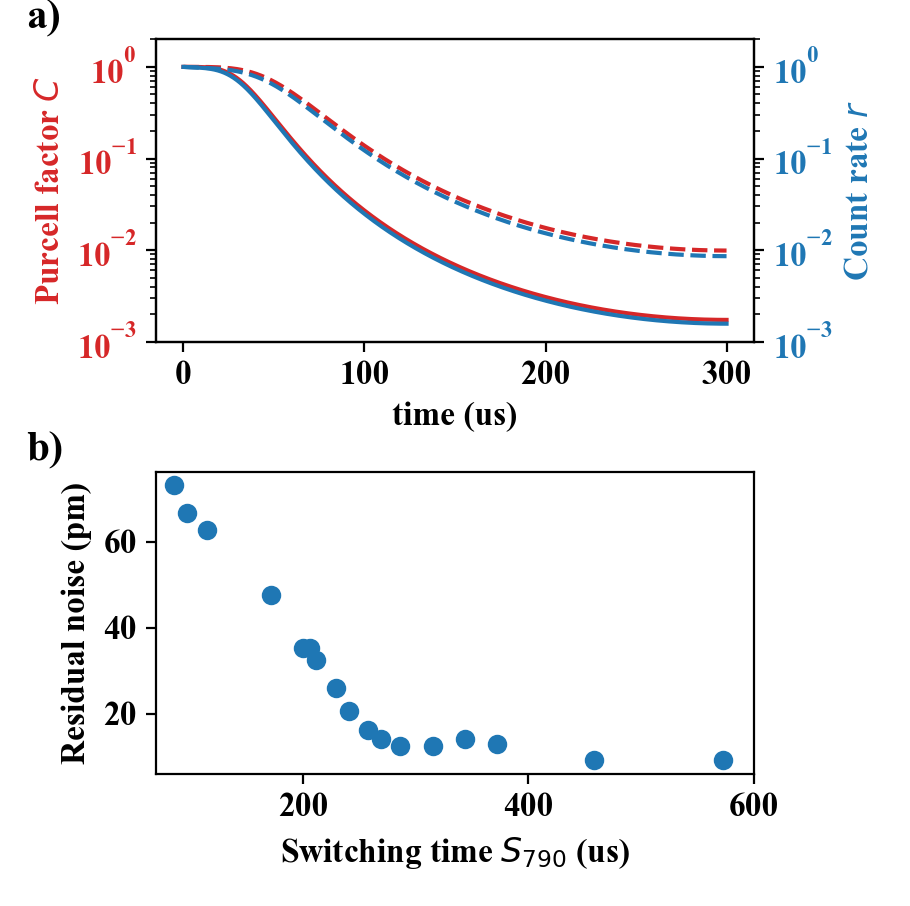}
\caption{a) Reduction of the Purcell factor and the countrate as function of time during $S_{790}=300~\mu$s. Solid lines is for the ideal detuning of $\Delta = \Delta_{790}/12$ and dashed line for $\Delta = \Delta_{790}/5$, the value extracted to match the data from Fig.~3e.
b) Residual noise as function of the switching time $S_{790}$}
\label{C_and_R_vs_time}
\end{figure}

For this experiment, we use $S_{790}=300~\mu$s. 
Assuming the cavity resonance at $1535~$nm is $\Delta_{1535}=\Delta_{790}/12$ and a Purcell factor $31$, we expect a total reduction of
$1/577$ in the Purcell factor and $1/634$ in the countrate  (see Fig.~SM.~\ref{C_and_R_vs_time}a).
Depending on the practical purpose a faster decoupling time might be required, which can be achieved at the expenses of a smaller reduction of the Purcell factor or the countrate. 
Here we define the switching time $S_{1535} $ as the time needed to decrease the countrate by a factor of 10.
For ideal parameters, we therefore expect $S_{1535} =67~\mu$s.

Now, we compare the expected reduction of the Purcell factor and the countrate with the measured values. Fig.~3e and 3f in the main text show the countrate as a function of time at the moment the cavity is detuned off- and on-resonance.
The solid line is the model defined in Eq. \ref{model}.
To have a good agreement with the data, we assumed an effective cavity linewidth 
$\Delta_{1535}=\Delta_{790}/5$, thus meaning that the effective total detuning given by $\Delta_{790}/\Delta$ is close to two times smaller than the expected.
The reason for the decreased effective detuning is the cavity stability, which is still not high enough to reach the maximum Purcell factor and the maximum switching time. 
For these parameters, the maximum reduction of the Purcell factor is $100$ and for the countrate is $116$.
In  Fig.~3e, the first vertical line indicates the beginning of the switching and the second is the time at which the countrate is reduced by a factor of 10, giving a $S_{1535}=106(10)~\mu$s.
In  Fig.~3f, the second vertical line indicates the end of the switching and the first is the time at which the countrate is a factor of 10 smaller than the maximum, for which the time interval is $120(10)~\mu$s.
Here, we see that by the time the cavity is back on resonance, the maximum countrate cannot be reach as population is lost during the switching process which occurs in a time scale comparable to the Purcell enhanced decay time.

In order to reduce further the switching time (assuming the cavity stability is not the limiting factor), one can simply move the piezo faster thus reducing $S_{790}$. 
However, a faster kick leads to increased residual mechanical noise. 
Fig.~SM.~\ref{C_and_R_vs_time}b shows the residual noise as function of the switching time $S_{790}$.
The residual noise is calculated as the standard deviation of the locking signal in the first $1~$ms after the cavity is set on resonance at $1535~$nm. 
As expected, the residual noise increases for a faster switching time $S_{790}$, that is, for faster switching frequency.
We estimate the switching frequency as $1 / (2 \times S_{790}) \approx 1.6~$kHz. 
The characteristic frequency of the residual noise is between $7-10~$kHz.
We attribute this residual motion to the fiber in the plane perpendicular to the cavity axis.
To first order, the cavity length should be robust against the lateral displacement of the fiber. 
However, a misalignment when placing the fiber in the positioner could lead to strong coupling between them, as confirmed in our experiment.
We see that while moving the fiber laterally, the cavity length is shifted, and we estimate the coupling to be in the $20-40\%$ range. 

In order to increase the switching time $S_{1535}$, we can follow several strategies:
\begin{itemize} 
\item perform a proper alignment of the fiber in order to minimize the coupling between the vibration modes,
\item increase the frequency of all mechanical eigenmodes,
\item increase the finesse of the $1535~$nm resonance,
\item use a faster growing function than $\sin^2$ or implement iterative learning algorithms to shape the signal sent to the piezo to minimize added noise
\end{itemize}
Altogether, we estimate that values of $S_{1535}$ in the microsecond scale could be achievable by combing these improvements if the cavity stability issue is addressed.

\section{multi exponential decay}\label{sec_mul}

\begin{figure}
\includegraphics[width=0.45\textwidth]{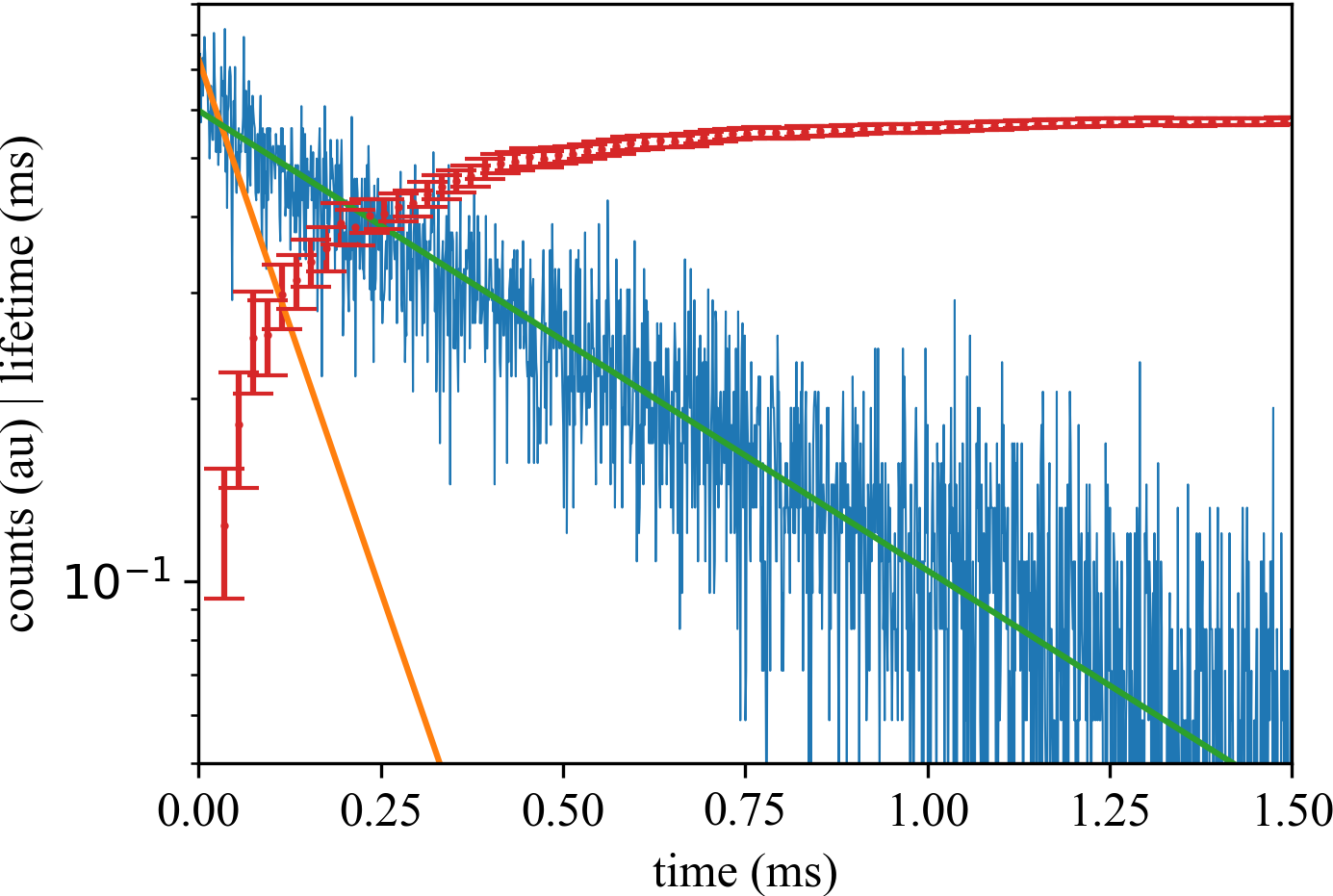}
\caption{
Blue: counts as function of the detection time $t$ for an input power of $11~\mu$W (constant background subtracted).
Red: lifetime $\tau_c$ obtained in a time window $[0-t]$.
The first red point corresponds to a time window $[0, 35]~\mu$s divided in 35 bins. 
Orange and green curves are the exponential decaying fits obtained for the first and last red points, giving values of $\tau_c=0.12(3)$ and $0.570(5)$ respectively.}
\label{sm_lt_vs_timebin}
\end{figure}

In order to study the presence of a multi exponential behavior in the Purcell enhanced fluorescence decay, we analyze the decay as function of the length of the detection time window. 
Fig.~SM.~\ref{sm_lt_vs_timebin}: blue shows a lifetime measurement performed over the same nanoparticle as the one shown in Fig.~2 but for longer acquisition time and for an input power of $11~\mu$W.
For this particular nanoparticle, the natural lifetime when it is not coupled to the optical cavity is measured to be $\tau_{n}=18(1)~$ms (see main text).
Red points in Fig.~SM.~\ref{sm_lt_vs_timebin} correspond to the extracted lifetime $\tau_c$ obtained in a time window $[0-t]$, the first of which corresponds to a time window $[0, 35]~\mu$s divided in 35 bins. 
A fit in the shortest time window reveals a lifetime $\tau_{c,\textrm{max}}=0.12(2)$~ms.
As the time window increases, the lifetime $\tau_c$ increases and it saturates for a time close to $1.25$~ms, giving a value of $\tau_{c,\textrm{avg}}=0.570(5)~$ms consistent with the value shown in the main text.
We then calculate the average and maximum measured Purcell factors
\begin{equation}
C= \frac{\tau_{n} }{\tau_{c}}-1
\end{equation}
yielding to $C_{\textrm{max}}=150(13)$ and $C_{\textrm{avg}}=30(1)$

The increase in the lifetime as function of the time window is an indication of multi-exponential decay.
The multi-exponential behavior can be  attributed to several factors.  
First, we consider the finite extension of the nanoparticle with respect of the cavity standing wave.
Assuming a nanoparticle sitting exactly in the maximum of the standing wave field, ions located far from the center of the nanoparticle will experience a smaller Purcell effect compared to those in the middle.
The Purcell factor will be then corrected by 
\begin{equation}
\epsilon_{\textrm{sw},d}= \cos^2(2 \pi ~d / \lambda), 
\end{equation}
where $d$ is the distance of a particular ion from the center of the nanoparticle.
In particular, for the studied nanoparticle with a diameter of $90~$nm, the maximum correction is $0.87$.
By assuming an spherical nanoparticle, we can then estimate an average correction of $\epsilon_{\textrm{sw}}=0.95$.

Second, the nanoparticle studied is compounded by several crystalline structures. 
Each of this structures have erbium ions located in a C2 symmetry crystallographic site thus 6 possible  $\bar d $ dipole orientations are possible. 
The Purcell factor for each ion in the nanoparticle is then corrected by 
\begin{equation}
 \epsilon_{\textrm{dip, }\hat d} = \cos^2  (\hat d \cdot \hat E) 
 \end{equation}
where $\bar E $ is the cavity electric field. 
As the number of sub crystalline structures is unknown, we assume no preferential direction for $\bar d$.
For an excitation pulse length of $10~\mu$s and for an input power of $11~\mu$W, most ions are already incoherently excited. 
Considering a constant excitation probability and that two out of three dipole orientations couple to the optical cavity, we can estimate an average correction of $\epsilon_{\textrm{dip}}=2/3$.
For this we assumed a constant collection efficiency $\beta=1$ as the collection efficiency stays above $90\%$ for a Purcell factor as low as $C=10$, and decreases to $1/2$ for $C=1$. 

Finally, we consider the fluctuation of the cavity resonance. 
We don't have a direct characterization for the data shown in Fig.~SM.~\ref{sm_lt_vs_timebin}.
However, typical normalized average transmission during the time the data was taken were between $0.4$ and $0.75$ measured for a cavity finesse of $6,000$ (performed with a bigger nanoparticle). 
Considering a mean value of $0.57$, we then extrapolate an average transmission of $0.22$ for a cavity with finesse of $16,000$ (as for the data discussed in this section).
We then introduce the correction factor 
\begin{equation}
\epsilon_{\textrm{disp}}=0.22,
\end{equation}
which takes into account that the Purcell factor is proportional to $\bar E^2$.

Now, the expected average Purcell factor can be calculated as 
\begin{equation}
C_{\textrm{exp}} \times ( \epsilon_{\textrm{sw}}  ~\epsilon_{\textrm{dip}} ~ \epsilon_{\textrm{disp}})=25,
\end{equation}
where $C_{\textrm{exp}}=176$ is the maximum expected Purcell factor (see main text). 
This value is slightly smaller than the measured one.
We attribute the discrepancy to a better cavity stability than the one assumed. 

In conclusions, by analyzing the Purcell factor at different time windows, we have seen that components as fast as 
$\tau_{c,\textrm{max}}=0.12(2)$~ms are present in the detected signal.
These ions experienced a Purcell factor of $C_{\textrm{max}}=150(13)$, which is close to the expected maximum value of $C_{\textrm{exp}}= 176$. 
In case the cavity stability issue is addressed, ions in the center of the nanoparticle and with the dipole moment aligned to the cavity electric field could be in principle isolated yielding to an up to five times higher signal to noise ratio, a level at which single ion detection would be feasible in our setup (see `Number of detected ions' section for a discussion of the signal to noise ratio).

\section{Number of detected ions}\label{sec_det_sig}

The probability $p_{det}$ to detect a photon generated in the cavity mode is given by 
\begin{equation}
p_{det} = \eta_{out} \times \eta_{mm} \times \eta_{col} \times \eta_{det} \times \eta_{g}
\end{equation}
where $\eta_{out}$ is the probability of the photon leaving trough the fiber as defined in Eq.~\ref{eta_out}, 
$\eta_{mm}$ is the mode matching between the fiber and the cavity mode~\cite{hunger2010_2},
$\eta_{col}$ is the collection path efficiency,
$\eta_{det}$ is the detector efficiency and $\eta_g$ is the proportion of the single photon in the detection time window.

\begin{figure}
\includegraphics[width=0.45\textwidth]{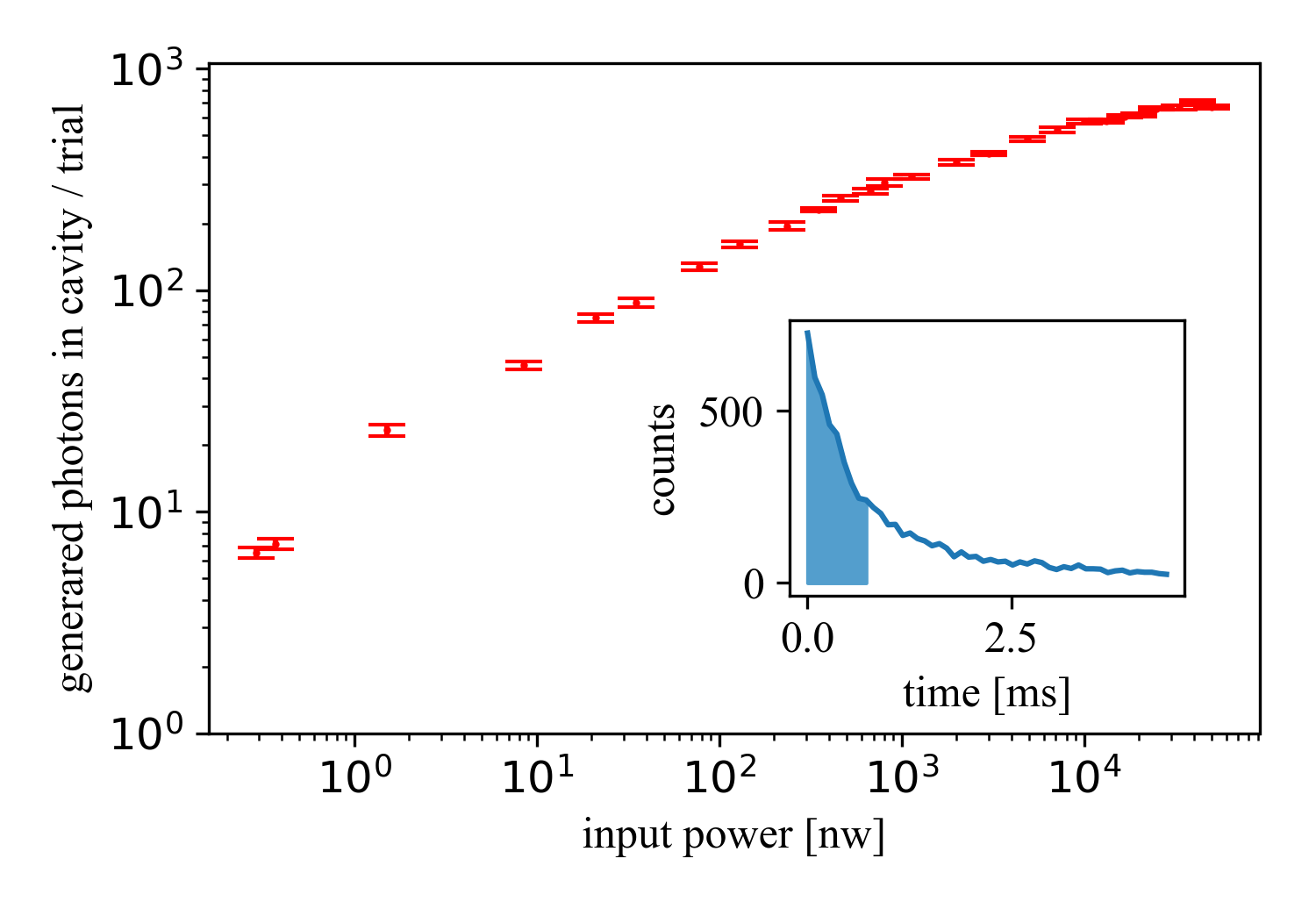}
\caption{
Number of photons generated in the cavity mode as a function of the excitation power before the cavity. 
Excitation pulse length is $500~\mu$s. 
The inset shows the detection window $[0- \tau_c]$, with $\tau_c=0.7~\mu$s.}
\label{sm_photons_vs_powerinput}
\end{figure}

For our cavity, $\eta_{out}=0.25$ when detecting via fiber and when the particle is aligned to the cavity mode, $\eta_{mm}$ is calculated to be $0.60$, $\eta_{col}$ is measured to be $0.3$, 
$\eta_{det}=0.1$, and $\eta_g=0.63$ for a detection time window $t_{det}=[0 - \tau_c]$ where $\tau_c$ is the lifetime of the emitter under Purcell enhancement.
All together then results in $p_{det}=0.28\%$.
Fig.~SM.~\ref{sm_photons_vs_powerinput} shows a measurements of the number of photons generated in the cavity as a function of the excitation input power at the input of the cavity.
For an input power of $7~$nW as in Fig.~2 (main text), the detected signal corresponds to $40$ intra cavity photons generated by $80$ ions (an ion is excited with almost 50\% probability, see Eq.~\ref{p_gen}).
For the first data point of Fig.~SM.~\ref{sm_photons_vs_powerinput} with input power of $330~$pW, the detected signal corresponds to $5$ intra cavity photons generated by $10$ ions.

In order to detect a single ion, we need to compare the probability of detecting a photon $p_{det}$ with the probability of detecting noise in the same time window $t_{det}$.
We~define the signal to noise ratio as 
\begin{equation}
S/N=\frac{p_{det}}{p_n},
\end{equation}
where $p_n$ is the background probability in the detection window $t_{det}$, in our case mostly due to the dark-count rate of the single photon detector (10 Hz).  
We~then calculate 
\begin{equation}
S/N=\frac{p_{det}}{\tau_c} \times \frac{1}{10~\textrm{Hz}}= 0.5. 
\end{equation}
This means that the number of photons that must be generated in the cavity to achieve a $S/N=1$ is equal 2. 
The probability $p_{gen}$  for an ion to emit a single photon in the cavity is indeed given by 
\begin{equation}
p_{gen}=p_{exc} \times \chi_{cav} \times \beta \approx 0.47,
\label{p_gen}
\end{equation}
where $p_{exc}$ is the excitation probability (0.5 since we excite incoherently). 
We therefore need to detect the fluorescence from four ions to achieve a $S/N=1$, which is currently not sufficient to reach high-fidelity detection of a single ion.  

Several solutions can be implemented to increase the signal in order to detect a single ion.
First, with the use of a superconducting nanowire single photon detector, which are specified to have detection efficiencies of close to 80\% (efficiency of our current detector is 10\%).
With this detector, the sensitivity could be increased by at least a factor of $8$, thus single photon detection would be already possible with no extra modification to the setup.  
In case the cavity stability is improved, ions that emit photons at a rate which is up to five times faster than the average can be in principle addressed (see 'Multi exponential decay' section), leading to an additional factor of $5$ in the signal to noise ratio. 
In addition, in our current cavity most of the light is emitted on the planar mirror side ($T_m = 2 T_f$). 
Future cavities with higher reflective mirrors will significantly increase the escape efficiency through the fiber. 
Even more, new mirrors will also lead to significantly higher cavity finesse and therefore increased Purcell enhancement.  
By increasing the reflectivity of the mirror in order to have a finesse of $40,000$, we expect that the cavity improvements will lead to a detection sensitivity improved by a factor $4$. 
We can therefore expect an increase in detection sensitivity by a factor 120 compared to the current sensitivity, which should allow us to reach high fidelity single photon detection.


%

\end{document}